\documentclass[useAMS,usenatbib]{mn2e}

\usepackage{graphicx}

\title[PKS 2254$-$367, a low redshift GPS radio source]{The multi-frequency parsec-scale structure of PKS 2254$-$367 (IC 1459): a luminosity-dependent break in morphology for the precursors of radio galaxies?}
\author[S.J. Tingay \& P.G. Edwards]{S.J. Tingay$^{1}$\thanks{E-mail:
s.tingay@curtin.edu.au} and P.G. Edwards$^{2}$\thanks{E-mail: Philip.Edwards@csiro.au}\\
$^{1}$International Centre for Radio Astronomy Research, Curtin University, Bentley, WA 6102, Australia\\
$^{2}$Australia Telescope National Facility, CSIRO Astronomy and Space Science, Epping, NSW 1710, Australia}
\begin{document}

\date{Submitted: 2014}

\pagerange{\pageref{firstpage}--\pageref{lastpage}} \pubyear{2014}

\maketitle

\label{firstpage}

\begin{abstract}
We present the first multi-frequency VLBI images of PKS 2254$-$367, a
Giga-hertz-Peaked Spectrum (GPS) radio source hosted by the nearby galaxy 
IC\,1459 ($D$=20.5\,Mpc).  PKS 2254$-$367 and the radio source in NGC\,1052
(PKS 0238$-$084; $D$=17.2\,Mpc) are the two closest GPS radio
sources to us, far closer than the next closest example, PKS
1718$-$649 ($D=59$\,Mpc).  As such, IC 1459 and NGC\,1052 offer
opportunities to study the details of the pc-scale radio sources as
well as the environments that the radio sources inhabit, across the
electromagnetic spectrum.  Given that some models for the origin and
evolution of GPS radio sources require a strong connection between the
radio source morphology and the gaseous nuclear environment, such
opportunities for detailed study are important.  Our VLBI images of
PKS 2254$-$367 show that the previously identified 
similarities between IC 1459 and NGC\,1052 continue onto the
pc-scale.  Both compact radio sources appear to have symmetric jets of
approximately the same luminosity, much lower than typically noted in
compact double GPS sources.  Similarities between PKS 2254$-$367 and
NGC\,1052, and differences with respect to other GPS galaxies, lead us to
speculate that a sub-class of GPS radio sources, with low luminosity
and with jet-dominated morphologies, exists and would be largely
absent from radio source surveys with $\sim$1 Jy flux density cutoffs.
We suggest that this possible low-luminosity, jet-dominated population
of GPS sources could be an analog of the FR-I radio galaxies, with the
higher luminosity lobe-dominated GPS sources being the analog of the
FR-II radio galaxies.
\end{abstract}


\section{Introduction}

Giga-hertz--Peaked Spectrum (GPS) sources represent at least 10\% cent of the
bright radio source population (O'Dea 1998)
and are thought to represent the early evolutionary stages of the
FR-II, and perhaps FR-I, radio galaxies (e.g., Snellen et al.\ 2003).
High angular resolution studies are therefore of
great importance for understanding the early stages of their evolution,
particularly for low-redshift GPS galaxies 
where high angular resolution corresponds to high linear resolution.
Multi-frequency observations additionally allow the spectral index
distribution to be studied: flat or inverted spectra identify regions of
active particle acceleration, whereas steeper spectra are consistent 
with an aging population of relativistic electrons (although these
interpretations can be complicated in high density regions by
the effects of absorption).

Multi-frequency, multi-epoch observations with the Australia Telescope
Compact Array (ATCA) (Tingay et~al.\ 2003a) established
PKS~B2254$-$367 as a GPS source.  In addition to the peaked radio
spectrum, PKS~B2254$-$367 displays all of the typical characteristics
of GPS sources, with compact structure seen on the maximum ATCA 6\,km
baselines, low fractional polarisation, and low variability over the
3.5 year period of the ATCA monitoring (Tingay, Edwards \& Tzioumis
2003b; Edwards \& Tingay 2004).  In general, GPS objects have marginal
variability, representing the variations due to the evolution of a jet
originating from a black hole and accretion disk system.  In GPS
objects, the contribution to the total flux density of the jet is
small in fractional terms, leading to low levels of variability. Long
term gradual flux density variability has been reported for OQ208 (Wu
et al.\ 2013).  PKS2254$-$367 has a more irregular variability,
similar to high frequency peaking GPS objects (e.g. \citet{or13} and
references therein).  Observations at 20\,GHz (Murphy et al.\ 2010)
and 95\,GHz (Sadler et al.\ 2008) confirm the PKS 2254$-$367 spectral
index steepens beyond 8.6\,GHz (the highest frequency of the original
ATCA observations).

Assuming a Hubble constant of 75 kms$^{-1}$Mpc$^{-1}$ and correcting
for the motion of our Galaxy in the direction of PKS 2254$-$367
relative to the Cosmic Microwave Background, the PKS 2254$-$367
redshift (0.006011, \cite{Zwa04}) indicates a distance of 20.5\,Mpc.
PKS~2254$-$367 is therefore one of the closest known GPS radio
sources.  NGC~1052 (z=0.005037) at 17.2\,Mpc and PKS~1718$-$649
(NGC~6328: 0.014428) at 59\,Mpc, in the same frame of reference, are
the only two other GPS radio sources within 100\,Mpc, for which the
detailed kinematics of the host galaxies can be studied (see, e.g.,
Labiano et al.\ 2007).  All three galaxies present strong evidence for
merger activity, an actively fuelled black hole, and high density
environments with which the radio sources interact (Tingay et
al.\ 2003b).

The host galaxy of PKS~B2254$-$367, IC~1459, is a giant elliptical
galaxy in a loose group of spirals.  IC~1459 is a LINER (Verdoes
Kleijn et~al.\ 2000), and has one of the strongest counter-rotating
core components of any observed elliptical, suggestive of merger
activity (Forbes, Franx, and Illingworth 1994).  The dust distribution
is irregular near the nucleus, indicating that infalling material may
currently be fuelling the active nucleus (Forbes et~al.\ 1994).  The
central black hole mass in IC 1459 is between
$4\times10^{8}$M$_{\sun}$ and $3\times10^{9}$M$_{\sun}$ (Cappellari
et~al.\ 2002).

\section{Observations and results}

Very Long Baseline Interferometry (VLBI) observations of PKS
2254$-$367 were made using the VLBA (excluding the northernmost
stations at Brewster, North Liberty and Hancock) at frequencies of
1.655 and 4.975 GHz on 2003 November 29 and at frequencies of 2.266
and 8.416 GHz on 2003 December 11.  Standard observing setups, with
two-bit quantisation, were used.

The 1.7 and 5.0\,GHz observations switched between these two bands on
a 15 minute timescale over the 6 hour duration of the observation.  A
32 MHz aggregate bandwidth was used at LCP.  The 2.3 and 8.4\,GHz
observations used the dual ``S/X'' band receiver to record RCP.
Switching between these frequencies was therefore not required, as
both are available simultaneously.  In this case, the aggregate
bandwidth per band is half of that which is available for the 1.7 and
5.0\,GHz observations, but the integration time is approximately
doubled.

The data were correlated on the VLBA processor and initial standard
data reduction was performed in AIPS, according to standard routines
(amplitude calibration, instrumental phase and delay calibration,
fringe-fitting etc).  The data were then exported to DIFMAP (Shepherd
1997) for editing, imaging, and modelfitting of the ($u,v$) data.

The aims of imaging the data at the maximum possible angular
resolution, to investigate the structure of the compact radio source,
and to obtain quantitative information on the spectral properties of
the source structure, are difficult to achieve given the poor ($u,v$)
coverage of the VLBA at southern declinations.  The absence of
Brewster, North Liberty, and Hancock from the array results in a large
hole in a largely east-west ($u,v$) coverage.  On the long baseline
side of the hole are baselines to the Mauna Kea and St Croix antennas.
On the short baseline side of the hole are baselines between the five
antennas concentrated in the south west US.

A consequence of the ($u,v$) coverage is that approximately matched
resolution images can be made at our four frequencies using a
combination of restrictions on the ($u,v$) data and weighting of the
visibilities (described in a sub-section below).  This approach
reveals useful qualitative information on the structure of the compact
radio source.  However, the uncertainties in deconvolution induced by
the poor ($u,v$) coverage do not allow quantitative information on the
spectral characteristics of the structure.

In order to best determine spectral information, the long baselines
need to be discarded and restrictions on the short baseline ($u,v$)
coverage are applied as a function of frequency to provide
visibilities matched in spatial frequency (without the need for
additional weighting) for component fitting in the ($u,v$) plane
(described in a sub-section below).

Using these two approaches, we determine limited information on the
structure and spectra of the compact radio source.

\subsection{Approximately matched resolution imaging}

In order to produce the highest angular resolution approximately
matched images of the compact structure, we started with the full
($u,v$) coverage at 1.7\,GHz.  At 2.3\,GHz, the same ($u,v$) coverage
was used but a Gaussian taper giving a weight of 0.5 at a ($u,v$)
distance of 20 M$\lambda$ (denoted 0.5, 20) was applied to produce an
image which approximately matched the resolution obtained at 1.7\,GHz.
At 5.0\,GHz, no Gaussian taper was used, but the long baselines to
Mauna Kea to Saint Croix were removed, to approximately match the
resolution at the two lower frequencies.  At 8.4\,GHz, the data on all
baselines to MK and SC were also removed from the dataset and a
Gaussian taper of 0.5, 30 was applied to the data, to approximately
match the resolution at the other frequencies.  All images were
produced using uniform weighting of the data, image sizes of 256
$\times$ 256 pixels, and pixel sizes of 1.0 mas.  Standard
deconvolution and self-calibration routines were used to form the
images.  As phase referencing was not used for these observations, the
self-calibration step in imaging (and indeed via fringe-fitting in
AIPS), forces the brightest component in the image at the centre of
the image.  As seen below, the brightest component is the same
component at all frequencies, thus the alignment of the images at the
different frequencies is secure to first order.  Small
frequency-dependent shifts in the centroid of the brightest component
are lost due to self-calibration.  However, a comparison of the
overall structure of the source allows a secure identification of the
different components of emission across the different frequencies.
These identifications are used to guide the model-fitting process
described in the next section.

Table~1 lists the tapers used at each frequency, the dimensions of the
resultant synthesised beams, and the image RMS values achieved.

\begin{table*}
\centering
\begin{minipage}{140mm}
\caption{Parameters of the images.  The beam parameters are semi minor axis FWHM $\times$ semi major axis FWHM @ beam position angle in degrees.  ($u,v$) coverage refers to the full VLBA (Full) or without the baselines to Mauna Kea ($-$MK) or Saint Croix ($-$SC).  Taper is described in the text.  RMS is the Root Mean Square in the residual image pixels.}
\begin{tabular}{@{}lcccc@{}}
\hline
Parameter                             & 1.7 GHz              & 2.3 GHz               & 5.0 GHz               & 8.4 GHz \\ \hline
Beam (mas$\times$mas @ PA $^{\circ}$) & 3.3$\times$9.8 @ 0.0 & 3.4$\times$10.3 @ 0.5 & 3.5$\times$11.0 @ 4.9 & 3.2$\times$11.8 @ 11.2  \\
($u,v$) coverage                      & Full                 & Full                  & $-$MK, $-$SC          & $-$MK, $-$SC \\
Taper (weight, uvdistance)            & NA                   & 0.5, 20               & NA                    & 0.5, 30 \\
RMS (mJy/beam)                        & 1.0                  & 1.1                   & 0.9                   & 0.8\\
\hline
\end{tabular}
\end{minipage}
\end{table*}

Finally, all images were restored with a common Gaussian restoring
beam of 3.5 mas x 11 mas at a PA of 6.0 degrees and are shown in
Figure~1.  This procedure follows the imaging techniques used by
\cite{tin01}.

\begin{figure*}
\includegraphics[width=150mm]{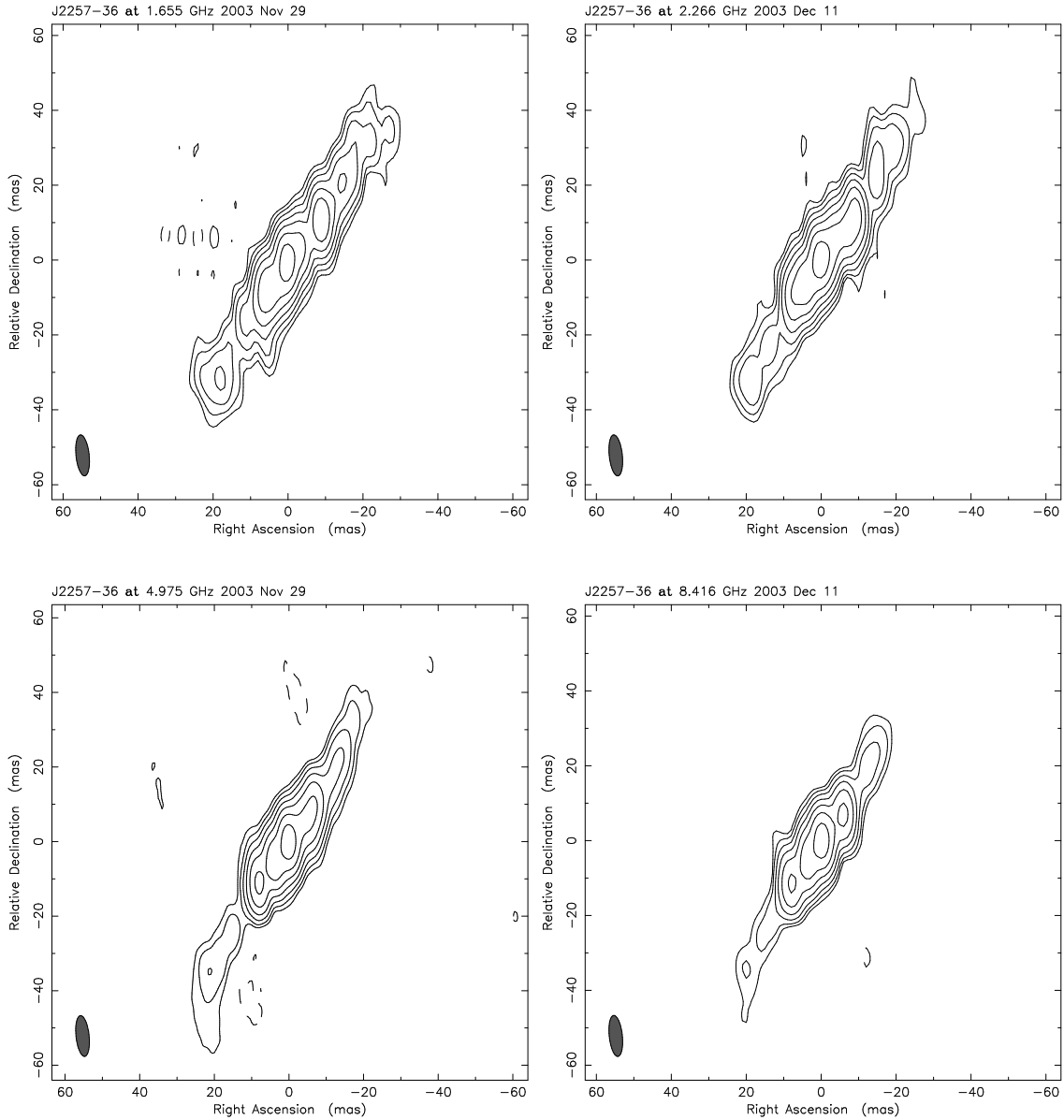} 
  \caption{VLBA images of PKS 2254$-$367.  All images have been
    restored with a beam of 11\,$\times$\,3.5\,mas at a position angle
    of 6$^\circ$. The image peaks are 243, 262, 251 and 240 mJy/beam,
    at 1.7, 2.3, 5.0 and 8.4\,GHz respectively. Contours are plotted
    at $-$1 (dashed), 1, 2, 4, 8, 16, 32, and 64\% of the image peak.
  }
\end{figure*}

\subsection{Matched resolution model-fitting}

As mentioned above, we seek to obtain limited spectral information on
the compact radio source by discarding the longest baselines to Mauna
Kea and St Croix and restricting the ($u,v$) coverages at all four
frequencies to a range of 0--9 M$\lambda$.  This allows the full short
baseline ($u,v$) coverage at 1.7\,GHz, and progressively more
restricted coverages at the higher frequencies.

The 1.7\,GHz dataset was fit with a model consisting of five circular
Gaussian components, using the task modelfit in DIFMAP.  We allowed
the flux densities, positions, and sizes of the five components to
vary until convergence.  This model was then fit to the data at the
three higher frequencies by keeping the positions and sizes of the
five components constant and allowing the flux densities to vary.
Good fits were found at all four frequencies using this method,
providing integrated flux densities for each component in order to
estimate the spectrum for each component.  Model component types other
than circular Gaussians were tested using this model-fitting approach
and were not found to provide significantly better fits to the data.
We justify the use of circular Gaussians as the simplest set of
components that provide a good fit to the data with the minimum number
of free parameters.

The errors on the VLBA flux density scale are approximately 5\%.  The
model fit results for all four frequencies are given in Table 2 and
the spectra are shown in Figure 2.

The results of the model-fitting are broadly consistent with the
results of the imaging shown in Figure 1, discussed below.

\begin{table*}
\centering
\begin{minipage}{140mm}
\caption{Flux densities of components fitted to ($u,v$) data in the range 0--9 M$\lambda$.  $S$ is the flux density (in Jy) of the model component at the frequency (in GHz) indicated by the subscript.  $r$ is the angular distance in milli-arcseconds of the component from the phase centre.  $\theta$ is the position angle in degrees (east of north) of the component.  $a$ is the Full Width at Half Maximum of a circular Gaussian component (in milli-arcseconds)}
\begin{tabular}{@{}cllll|rrr@{}}
\hline
Component & $S_{1.7}$     & $S_{2.3}$     & $S_{5.0}$     & $S_{8.6}$     & $r$  & $\theta$ &  $a$ \\ \hline
A         & 0.49$\pm$0.03 & 0.58$\pm$0.03 & 0.53$\pm$0.03 & 0.48$\pm$0.03 &  1.6 &   $-$174 &  0.0 \\
B         & 0.33$\pm$0.02 & 0.30$\pm$0.02 & 0.16$\pm$0.01 & 0.10$\pm$0.01 & 15.2 &    $-$42 &  7.3 \\
C         & 0.17$\pm$0.02 & 0.19$\pm$0.02 & 0.17$\pm$0.02 & 0.16$\pm$0.02 & 16.2 &      149 &  6.9 \\
D         & 0.08$\pm$0.01 & 0.04$\pm$0.01 &       $<$0.03 &       $<$0.03 & 40.9 &    $-$29 & 10.6 \\
E         & 0.04$\pm$0.01 & 0.05$\pm$0.01 & 0.04$\pm$0.01 & 0.05$\pm$0.01 & 42.0 &      153 &  6.8 \\
\hline
\end{tabular}
\end{minipage}
\end{table*}

\section{Discussion}

The analysis above provides information on the structure and spectral
properties of PKS 2254$-$367.  As one of only a small number of GPS
radio sources within 100\,Mpc, it is important to determine the basic
properties of the source and compare them to the other nearby, and
more distant, GPS sources.

A point of interest is how the nearby GPS sources, such as PKS
2254$-$367, relate to the so-called compact symmetric objects (CSOs).
CSOs are compact ($<$1 kpc in projected extent) and have highly
symmetric structures (Wilkinson et al.\ 1994); they are thought to be
the progenitors of Fanaroff-Riley type II radio galaxies (Readhead et
al.\ 1996).  Snellen et al.\ (2004) show that a very high percentage
of low redshift ($0.005 < z < 0.16$), compact ($< 2''$), and
relatively weak ($> 100$ mJy at 1.4\,GHz) radio sources associated
with galaxies bright at optical wavelengths can be classified as GPS
or CSS (compact steep spectrum) radio sources, and are, therefore,
perhaps young radio galaxies.

We therefore examine PKS 2254$-$367 to determine how it relates to
the classes of objects (GPS, CSS, CSO) thought to be the precursors of
powerful radio galaxies, and in particular other nearby radio sources
in these categories.

Tingay et al.\ (2003b) identified PKS 2254$-$367 ($D$=20.5\,Mpc) as
one of three low redshift GPS radio galaxies, along with PKS
1718$-$649 ($D$=59\,Mpc) and NGC\,1052 ($D$=17.2\,Mpc).  PKS
1718$-$649 has a so-called classical double morphology.  Tingay et
al.\ (2003b) listed a number of similarities between NGC\,1052 and the
host galaxy of PKS 2254$-$267 (IC 1459).  Both galxies have
counter-rotating cores (Forbes et al.\ 1994; Bettoni et al.\ 2001) and
LINER spectra (Verdoes Kleijn et al.\ 2000; Gabel et al.\ 2000).  Both
radio sources are predominantly compact and have peaked spectra (de
Vries, Barthel, \& O'Dea 1997; Tingay et al.\ 2003b).  All three low
redshift GPS sources support models in which a dense nuclear
environment and merger activity are the cause for the mostly compact
radio structure.  PKS 2254$-$367 and NGC\,1052 are hosted in cluster
dominant (CD) elliptical galaxies.

Of the three bright components that dominate the pc-scale structure of
PKS 2254$-$367, those labelled A, B, and C in Table~1, A has a
marginally peaked spectrum, sits at the centre of the structure, is
the brightest component, and is the most compact of the components.
We therefore identify component A as likely to correspond to the
nucleus of the radio source, presumably coincident with the
supermassive black hole known to reside in the galaxy.

The pc-scale radio source therefore appears to be double-sided, with
jet-like extensions comprised of the bright inner component B and the
weaker outer component D towards the north-west, and the bright inner
component C and the weaker outer component E towards the south-east.
This structure is reflected in both the modelfitting results and the
imaging results.  The north-west components (B and D) have spectra
that steepen above 2.3\,GHz.  The south-east components (C and E) have
flat spectra.  The spectra for the five components are shown in
Figure~2.

\begin{figure*}
\includegraphics[angle=-90,origin=c,width=150mm]{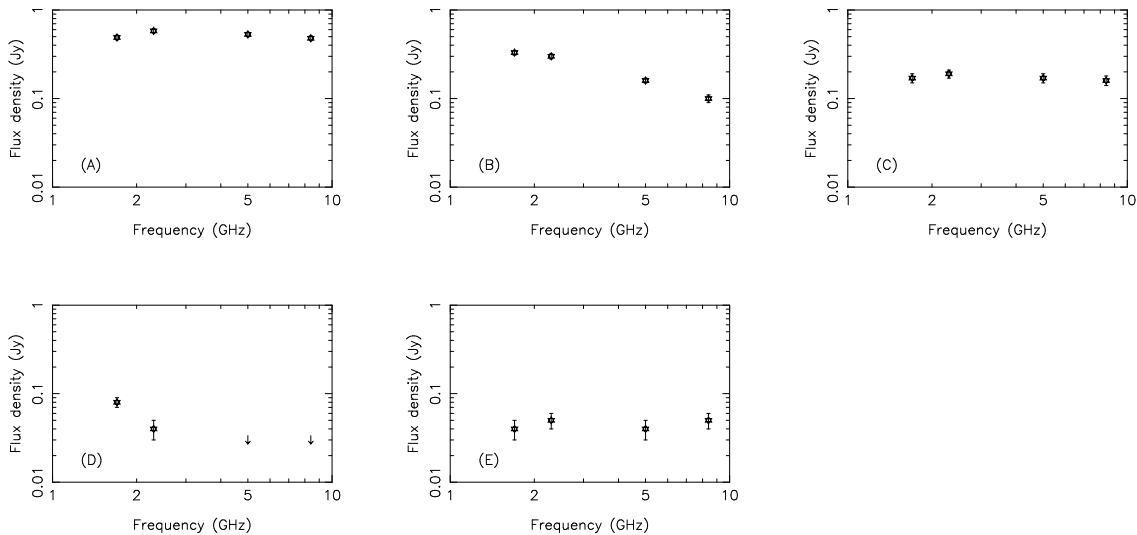} 
  \caption{Spectra for the modelfit components of Table 2.}
\end{figure*}

PKS 2254$-$367 therefore seems to be a candidate CSO, with a symmetric
structure centred on component A.  In comparison to the CSO candidates
found in the COINS survey (Peck and Taylor 2000), PKS 2254$-$367 seems
most similar to the sources J0427$+$41 and J1546$+$00.  Both these
sources appear to have a strong core (more dominant than in PKS
2254$-$367, indicating that these cores may be beamed) and double
sided jets.  However, such sources make up a small minority of the
candidates found by Peck and Taylor (2000).  \cite{or04} report VLBA
images of a sub-sample of CSS objects and out of 18 objects (from a
parent sample of 87 objects), two have morphologies similar to PKS
2254$-$367, possibly without as strong a central core component
(B3-VLA 1242$+$410 and B3-VLA 2358$+$406).  Most candidate CSOs are
dominated by strong edge-brightened lobes and have weak or absent jets
and weak or absent cores.  The more frequently observed morphology in
the Peck and Taylor sample is more reminiscent of PKS 1718$-$649.

Our new VLBA data for PKS 2254$-$367 shows that similarities between
PKS 2254$-$367 and NGC\,1052 continue to be seen for the pc-scale radio
source.

From comprehensive multi-epoch VLBI monitoring, NGC\,1052 has been
revealed to possess bi-directional jets with apparent speeds of up to
$\sim$0.38$c$ that appear to be orientated in the plane of the sky
(Vermeulen et al.\ 2003; Lister et al.\ 2013).  The jets in PKS
2254$-$367 also appear to be bi-directional, with a high degree of
symmetry, although similar multi-epoch VLBI observations would be
required to unambiguously demonstrate this through the detection of
motion in the jet components.  With a spacing between observing epochs
of only 12 days, obviously no motion has been detected in PKS
2254$-$367 but the opposing jets appear highly symmetric with regard
to their surface brightnesses, indicating that they are in the plane
of the sky and/or not significantly relativistic.

The largest extent of the jets in NGC\,1052 is approximately 3\,pc
(Vermeulen et al.\ 2003).  In PKS 2254$-$367, the jets have a
projected extent of approximately 8\,pc.

The integrated monochromatic luminosity at 5\,GHz of PKS 2254$-$367 is
$\sim 7 \times 10^{22}$\,W\,Hz$^{-1}$, and the compact radio source in
NGC\,1052 has a very similar luminosity, $\sim5 \times
10^{22}$\,W\,Hz$^{-1}$.  PKS 1718$-$649 has a classical double
morphology with an extent of 2 pc and a monochromatic luminosity at
5\,GHz of $\sim 2 \times 10^{24}$\,W\,Hz$^{-1}$ (Tingay et al.\ 1997).
Unlike NGC\,1052, no motion in the structure of PKS 1718$-$649 has
been detected following multi-epoch VLBI observations, with an
estimated upper limit on the separation speed of the double components
of 0.08\,$c$ (Tingay et al.\ 2002).

In NGC\,1052, Vermeulen et al.\ (2003) found strong evidence for
free-free absorption within the central pc.  We cannot examine this
possibility for PKS 2254$-$367, since the limited frequency range,
($u,v$) coverage, and resolution of our data, compared to the
NGC\,1052 data, make detection of free-free absorption effects
difficult.  In general, distinguishing between free-free absorption
and synchrotron self-absorption is difficult in GPS/CSO objects, as
examined in detail by Tingay and de Kool (2003) for PKS 1718$-$649.
More recent work on the absorption mechanism in PKS 1718$-$649 by
Tingay et al.\ (2015) shows that free-free absorption is likely to be
the cause of its spectral peak at $\sim$3\,GHz.

The similarities between NGC\,1052 and PKS 2254$-$367 on the pc-scale
are interesting, as are the differences between these two sources and
PKS 1718$-$649 and similar lobe-dominated GPS sources.

With the new images of PKS 2254$-$367, the two closest GPS sources are
of low luminosity and have extended low-power jets (subluminal for
NGC\,1052, unknown for PKS 2254$-$367).  The next closest sources are
substantially more luminous -- fifty times so in the case of PKS
1718$-$649 -- and are dominated by powerful double morphologies
(sometimes with weak cores and symmetric jets).  PKS 2254$-$367 and
NGC\,1052 are below the lower edge of the distribution of luminosities
of the sample of low redshift GPS and CSS sources of Snellen et
al.\ (2004), as all sources in that survey are more luminous than
$\sim 10^{23}$\,W\,Hz$^{-1}$ at 5\,GHz.  The objects with somewhat
similar morphologies to PKS 2254$-$367 from \cite{or04} have
luminosities substantially higher, of order
$10^{27}$\,W\,Hz$^{-1}$\,$h^{-1}$ at 400\,MHz (these are CSS objects,
with spectra still rising at 400 MHz).

We speculate that objects such as NGC\,1052 and PKS 2254$-$367
represent a sub-class of GPS radio source which are compact, of low
luminosity, and jet dominated.  They are differentiated from many
other GPS sources (such as PKS 1718$-$649) by the lack of powerful,
compact lobes.

Such a possibility recalls the Fanaroff-Riley lumonosity$-$morphology
break seen in large-scale radio galaxies (Fanaroff \& Riley 1974).
The low luminosity radio galaxies (FR-I type) have prominent
(two-sided or sometimes one-sided) jets and centre-brightened or
indistinct radio lobes.  The great majority of the higher luminosity
radio galaxies (FR-II type) are dominated by edge-brightened lobes.
They typically have apparently weak cores and weak jets.  The apparent
weakness of the jets is ascribed to relativistic beaming along the jet
direction.

NGC\,1052 and PKS 2254$-$367 may be the low-luminosity, jet-dominated
analogs of FR-I type radio galaxies.  Objects like PKS 1718$-$649 may
be the higher luminosity analog of FR-II type radio galaxies.  A
possible implication of such a luminosity$-$morphology break could be
that the origin of the difference between FR-I and FR-II type radio
galaxies lies on the sub-parsec-scale, where the jet is formed close
to the black hole and accretion disk.  Such schemes are favoured in
the theoretical models of Meier et al.\ (1997), who suggest that the
FR-I/FR-II break is due to the magnetic field strength near the black
hole.

However, other explanations may exist for the morphology of NGC\,1052
and PKS 2254$-$367.  The nuclear environment is thought to play a
major role in the appearance and evolution of GPS radio sources, and
this seems likely in the case of these two objects.  In this case, the
sub-pc environment of the black hole and accretion disk system would
not be of primary relevance.  Support for this view can be found in
the fact that the properties of the pc-scale jets in FR-I and FR-II
objects, as seen with VLBI, are indistinguishable in terms of
morphology and apparent speed, for example as reported by
\cite{gio01}.

Very few sources like PKS 2254$-$367 or NGC\,1052 have been studied in
detail since even doubling their distance to a modest $\sim40$\,Mpc
would mean that they lie a factor of four below the 1\,Jy cutoff that
is typical of previous radio source surveys.  Multi-frequency surveys
of weak radio sources would be required to find such objects.  Beyond
$\sim 50$\,Mpc only the higher luminosity compact double and GPS
quasars (relativistically beamed emission) could expect to be
detected.  Even surveys that aim to improve our knowledge of the less
luminous GPS source population, such as those by Snellen et
al.\ (2004) or Peck and Taylor (2000), would struggle to detect PKS
2254$-$367 or NGC\,1052 at distances $>$50\,Mpc.

With only PKS 2254$-$367 and NGC\,1052 to suggest the possibility of a
low-luminosity, jet dominated population of GPS sources which is the
analog of FR-I radio galaxies, it is important to collect further data
for other nearby candidate objects.  The information on NGC\,1052 and
PKS 2254$-$367 suggest a way forward.  We should target a sample of
weak (10--100 mJy corresponding to luminosities of 
$10^{21 - 22}$\,W\,Hz$^{-1}$ at distances of $\sim$10 -- $\sim$70\,Mpc) 
radio sources in nearby elliptical galaxies.  Such radio sources have 
been examined with rudimentary high resolution observations previously
(e.g., Slee et al.\ 1994) but without examination of broad-band radio
spectra (to find peaked spectra) or detailed imaging.

Some progress in this direction has been made by de Vries et
al.\ (2009), who report VLBI observations of sources in the CORALZ
sample of Snellen et al.\ (2004).  An interesting example is
J105731$+$405646 with a 1.4 GHz flux density of 47\,mJy and a clearly
peaked spectrum near 1.3 GHz.  Snellen et al.\ give a redshift of
0.008, with a corresponding luminosity of $L=10^{21.59}$\,W\,Hz$^{-1}$
at 5\,GHz, however a redshift of 0.025 is also given in the literature
for this object (Miller et al.\ 2002).  In any case, de Vries et
al.\ find the object is unresolved in their EVN + VLBA observation,
with an image peak of 14 mJy/beam at 5\,GHz.

\section{Conclusions}

The nuclear radio sources in IC 1459 and NGC\,1052 are remarkably
similar, as are the properties of the host galaxies themselves.  Both
compact radio sources appear to have symmetric jets of approximately
the same luminosity, much lower than typically noted in compact double
GPS sources.  Similarities between PKS 2254$-$367 and NGC\,1052 and
differences with respect to other GPS galaxies lead us to speculate
that a sub-class of GPS radio sources, with low luminosity and with
jet-dominated morphologies, exists and would be largely absent from
previous radio source surveys.  We suggest that a low-luminosity,
jet-dominated population of GPS sources may be the analog of FR-I
radio galaxies, with the higher luminosity lobe-dominated GPS sources
being the analog of FR-II radio galaxies.

\section*{Acknowledgments}
We thank an anonymous referee for very useful comments on this
manuscript that prompted significant improvements.  The National Radio
Astronomy Observatory is a facility of the National Science Foundation
operated under cooperative agreement by Associated Universities, Inc.
SJT acknowledges the support of the Western Australian State
government, in the form of a Western Australia Premier’s Fellowship.

\label{lastpage}
\end{document}